\documentclass[a4paper,UKenglish,cleveref, autoref]{lipics-v2019}

\usepackage{algorithm}
\usepackage{microtype}
\usepackage{amssymb}
\usepackage{amsmath}
\usepackage{amsthm}
\usepackage{setspace} 
\usepackage{bm}
\usepackage{times}
\usepackage{fancybox}
\usepackage{color}
\usepackage{latexsym}
\usepackage{verbatim}

\newtheorem{observation}{Observation}[theorem]

\newcommand{\pit}{\mbox{\rm PIT}}

\newcommand{\w}{\mbox{\rm W}}

\newcommand{\kmmd}{{\mbox{\rm{\textit{k}{-}M\scriptsize{M}\normalsize{D}}}}}

\newcommand{\kMLC}{{\mbox{\rm{(\textit{k,n}){-}M\scriptsize{L}\normalsize{C}}}}}

\usepackage{thm-restate}
\usepackage{hyperref}

\newcommand{\ABP}{\mbox{\small\rm ABP}}
\newcommand{\e}{\boldsymbol{e}}

\newcommand{\F}{\mathbb{F}}

\newcommand{\C}{\mathbb{C}}
\newcommand{\Z}{\mathbb{Z}}

\newcommand{\Q}{\mathbb{Q}}

\renewcommand{\angle}[1]{{\langle} #1 {\rangle}}

\DeclareMathOperator{\poly}{\mbox{\small\rm poly}} 
  
 \DeclareMathOperator{\perm}{\mbox{\small\rm Perm}}

  \DeclareMathOperator{\ktreenew}{\mbox{\rm \textit{k}-Tree}} 
 \DeclareMathOperator{\tdominatingset}{\mbox{\rm \textit{t}-Dominating Set}} 
 \DeclareMathOperator{\mdimensionalkmatching}{\mbox{\rm \textit{m}-Dimensional \textit{k}-Matching}} 

 \DeclareMathOperator{\rper}{\mbox{\rm rPer}}

\newenvironment{appendix-lemma}[1]{\vspace{0.1in}\noindent{\bf Lemma~#1~} \em }{\vspace{0.1in}}
\newenvironment{appendix-theorem}[1]{\vspace{0.1in}\noindent{\bf Theorem~#1~} \em }{\vspace{0.1in}}

\usepackage{microtype}

\title{Fast Exact Algorithms Using Hadamard Product of Polynomials}

\titlerunning{Fast Exact Algorithms Using Hadamard Product of Polynomials}

\author{V. Arvind}{Institute of Mathematical Sciences (HBNI), Chennai, India}{email: arvind@imsc.res.in}{}{} 
\author{Abhranil Chatterjee}{Institute of Mathematical Sciences (HBNI), Chennai, India}{email: abhranilc@imsc.res.in}{}{} 
\author {Rajit Datta}{Chennai Mathematical Institute, Chennai, India}{email: rajit@cmi.ac.in}{}{}
\author {Partha Mukhopadhyay}{Chennai Mathematical Institute, Chennai, India}{email: partham@cmi.ac.in}{}{}

\authorrunning{V.Arvind, A.Chatterjee, R.Datta and P.Mukhopadhyay} 

\Copyright{V. Arvind and Abhranil Chatterjee and Rajit Datta and Partha Mukhopadhyay}



\supplement{}

\nolinenumbers 

\hideLIPIcs  

\begin{document}

\maketitle

\begin{abstract}
Let $C$ be an arithmetic circuit of $\poly(n)$ size given as input
that computes a polynomial $f\in\F[X]$, where
$X=\{x_1,x_2,\ldots,x_n\}$ and $\F$ is any field where the field
arithmetic can be performed efficiently. We obtain new algorithms for
the following two problems first studied by Koutis and Williams
\cite{Kou08, Wi09, KW16}.

$\kMLC$: Compute the sum of the coefficients of all degree-$k$
multilinear monomials in the polynomial $f$. 

$\kmmd$: Test if there is a nonzero degree-$k$ multilinear monomial in
the polynomial $f$.

Our algorithms are based on the fact that the Hadamard product $f\circ
S_{n,k}$, is the degree-$k$ multilinear part of $f$, where $S_{n,k}$
is the $k^{th}$ elementary symmetric polynomial.

\begin{itemize}

\item For $\kMLC$ problem, we give a deterministic algorithm of run
  time $O^*(n^{k/2+c\log k})$ (where $c$ is a constant), answering an
  open question of Koutis and Williams \cite[ICALP'09]{KW16}.  As
  corollaries, we show $O^*(\binom{n}{\downarrow k/2})$-time exact
  counting algorithms for several combinatorial problems: $\ktreenew,
  \tdominatingset, \mdimensionalkmatching$.

\item For $\kmmd$ problem, we give a randomized algorithm of run time
  $4.32^k\cdot\poly(n,k)$. Our algorithm uses only $\poly(n,k)$
  space. This matches the run time of a recent algorithm \cite{BDH18}
  for $\kmmd$ which requires exponential (in $k$) space.
 \end{itemize}
  Other results include fast deterministic algorithms for $\kMLC$ and
  $\kmmd$ problems for depth three circuits.
\end{abstract}

\newpage
\section{Introduction}\label{intro}

Koutis and Williams \cite{ Kou08, Wi09, KW16} introduced and studied
two algorithmic problems on arithmetic circuits. Given as input an
arithmetic circuit $C$ of $\poly(n)$ size computing a polynomial
$f\in\F[x_1,x_2,\ldots,x_n]$, the $\kMLC$ problem is to compute the
sum of the coefficients of all degree-$k$ multilinear monomials in the
polynomial $f$, and the $\kmmd$ problem is to test if $f$ has a
nonzero degree-$k$ multilinear monomial.

These problems are natural generalizations of the well-studied
$k$-path detection and counting problems in a given graph \cite{Kou08}
as well as several other combinatorial problems like $\ktreenew$,
$\tdominatingset$, $\mdimensionalkmatching$ \cite{KW16}, well-studied
in the parameterized complexity, reduce to these problems. In fact,
the first randomized FPT algorithms for the decision version of these
combinatorial problems were obtained from an $O^*(2^k)$~\footnote{The
  $O^*$ notation suppresses polynomial factors.}  algorithm for
$\kmmd$ for monotone circuits using group algebras~\cite{Kou08,
  Wi09,KW16}. Recently, Brand et al. \cite{BDH18} have given the first
randomized FPT algorithm for $\kmmd$ for general circuits that runs in
time $O^*(4.32^k)$. Their method is based on exterior algebra and
color coding~\cite{AYZ95}.

In general, the exact counting versions of these problems are
$\#\w[1]$-hard. For these counting problems, improvements to the
trivial $O^*(n^k)$ time exhaustive search algorithm are known only in
some cases (like counting $k$-paths)~\cite{BHKK09}. Since an
improvement for $\kMLC$ over exhaustive search will yield faster exact
counting algorithms for all these problems, Koutis and Williams
\cite{KW16} pose this as an interesting open problem. They give an
algorithm of run time $O^*(n^{k/2})$ to compute the \emph{parity} of
the sum of coefficients of degree-$k$ multilinear monomials.

The techniques based on group algebra \cite{ Kou08, KW16} and exterior
algebra \cite{BDH18} can be broadly classified as \emph{multilinear
  algebra} techniques.  We give a new approach to the $\kmmd$, $\kMLC$
problems, and related problems. Our algorithm is based on computing
the \emph{Hadamard product of polynomials}.  The Hadamard product
(also known as Schur product) generally refers to Hadamard product of
matrices and is used in matrix analysis. We consider the Hadamard
product of polynomials (e.g., see \cite{AJS09}). Given polynomials
$f,g\in\F[X]$, their Hadamard product is defined as $ f\circ g =
\sum_{m} ([m]f \cdot [m]g) m, $ where $[m]f$ denotes the coefficient
of monomial $m$ in $f$.

The Hadamard product is a useful tool in \emph{noncommutative}
computation~\cite{AJS09, AS18}.  A contribution of the present paper
is to develop an efficient way to implement Hadamard product in the
\emph{commutative} setting which is useful for designing FPT and exact
algorithms. As mentioned above, the Hadamard product has been useful
in arithmetic circuit complexity results, e.g., showing hardness of
the noncommutative determinant \cite{AS18}. Transferring techniques
from circuit complexity to algorithm design is an exciting area of
research. We refer the reader to the survey article of Williams
\cite{rw14}, see also \cite{rw14b}.
\vspace{0.2 cm}

\noindent\textbf{This paper.}  We apply the Hadamard product of
polynomials in the setting of commutative computation. This is
achieved by combining earlier ideas~\cite{AJS09, AS18} with a
symmetrization trick shown in Section~\ref{section-framework}. We then
use it to design efficient algorithms for $\kMLC$, $\kmmd$ and related
problems.

Consider the elementary symmetric polynomial $S_{n,k}$ of degree $k$
over the $n$ variables $x_1,x_2,\ldots,x_n$. By definition, $S_{n,k}$
is the sum of all the degree-$k$ multilinear monomials.  Computing the
Hadamard product of $S_{n,k}$ and a polynomial $f$ \emph{sieves out}
precisely the degree-$k$ multilinear part of $f$. This connection with
the symmetric polynomial gives the following result.

\begin{restatable}[]{thm}{kmlc}\label{kmlc-thm}
The $\kMLC$ problem for any arithmetic circuit $C$ of $\poly(n)$ size,
has a deterministic $O^*(n^{k/2+c\log k})$ time algorithm where $c$ is
a constant.
\end{restatable}

The field $\F$ could be any field where the field operations can be
efficiently computable. The above run time $O^*({n^{k/2 + c \log k}})$
(where $c$ is a constant) beats the naive $O^*({n^k})$ bound,
answering the question asked by Koutis and Williams~\cite{KW16}.

An ingredient of the proof is a result in \cite{AS18} that allows us
to efficiently compute the Hadamard product of a noncommutative
algebraic branching program ($\ABP$) with a noncommutative polynomial
$f$, even with only black-box access to $f$ that allows evaluating $f$
on matrix-valued inputs. The other ingredient is an algorithm of
Bj\"{o}rklund et al.~\cite{BHKK10} for evaluating rectangular
permanent over noncommutative rings, that can be viewed as an
algorithm for evaluating $S^*_{n,k}$ (a symmetrized noncommutative
version of $S_{n,k}$) over matrices.  Now, applying the routine
conversion of a commutative circuit to an $\ABP$, which incurs only a
quasi-polynomial blow-up, we get a faster algorithm for $\kMLC$ of
general circuits.  As applications of Theorem \ref{kmlc-thm} we obtain
improved counting algorithms for $\ktreenew, \tdominatingset$, and
$\mdimensionalkmatching$.

The next algorithmic result we obtain is the following.

\begin{restatable}[]{thm}{kmmd-thm}\label{kmmd-thm}
  The $\kmmd$ problem for any arithmetic circuit $C$ of $\poly(n)$
  size, has a randomized $O^*(4.32^k)$ time and polynomial
  space-bounded algorithm.
\end{restatable}

Again, the field $\F$ could be any field where the field operation can
be efficiently computable.  We briefly sketch the proof idea. Suppose
that $C$ is the input arithmetic circuit computing a homogeneous
polynomial $f$ of degree $k$. We essentially show that $\kmmd$ is
reducible to checking if the Hadamard product $f\circ C'$ is nonzero
for some circuit $C'$ from a collection of homogeneous degree-$k$
depth two circuits. This collection of depth two circuits arises from
the application of color coding \cite{AYZ95}. Furthermore, the
commutative Hadamard product $f\circ C'$ turns out to be computable in
$O^*(2^k)$ time by a symmetrization trick combined with Ryser's
formula for the permanent.  The overall running time (because of
trying several choices for $C'$) turns out to be
$O^*(4.32^k)$. Finally, checking if $f\circ C'$ is nonzero reduces to
an instance of polynomial identity testing which can be solved in
randomized polynomial time using Demillo-Lipton-Schwartz-Zippel
Lemma~\cite{DL78, Zip79, Sch80}.  The technique based on Hadamard
product seems to be quite different than the exterior algebra based
technique. Another difference is that, our algorithm uses $\poly(n,k)$
space whereas the algorithm in \cite{BDH18} takes exponential space.


Next, we state the results showing fast deterministic algorithms for
depth-three circuits. We use the notation
$\Sigma^{[s]}\Pi^{[k]}\Sigma$ to denote depth three circuits of top
$\Sigma$ gate fan-in $s$ and the $\Pi$ gates compute the product of
$k$ homogeneous linear forms over $X$.

\begin{restatable}[]{thm}{kmmddepththree} \label{depth3-kmmd}
Given any homogeneous depth three $\Sigma^{[s]}\Pi^{[k]}\Sigma$
circuit of degree $k$, the $\kMLC$ problem can be solved in
deterministic $O^{*}(2^k)$ time. Over $\Z$, the $\kmmd$ problem can be
solved in deterministic $O^{*}(4^k)$ time. Over finite fields, $\kmmd$
problem can be solved in deterministic $e^k k^{O(\log k)}(2^{ck} +
2^k)\cdot\poly(n,k,s)$ time, where $c\leq 5$.
\end{restatable}

Here the key observation is that we can efficiently compute the
commutative Hadamard product of a depth three circuit with \emph{any}
circuit. It is well-known that the elementary symmetric polynomial
$S_{n,k}$ can be computed using an algebraic branching program of size
$\poly(n,k)$.

We compute the Hadamard product of the given depth three circuit with
that homogeneous branching program for $S_{n,k}$, and check whether
the resulting depth three circuit is identically zero or not. The same
idea yields the algorithm to compute the sum of the coefficients of the
multilinear terms as well.

\vspace{0.2 cm}
\noindent\textbf{Related Work.} Soon after the first version of our
paper~\cite{ACDM18} appeared in ArXiv, an independent work 
\cite[v1]{Pra18} \footnote{See the final version \cite{Pra18} to be appeared in
  FOCS 2019.} also considers the $\kmmd$ and $\kMLC$ problems. The
main ingredient of \cite{Pra18} is the application of a nontrivial
Waring decomposition over rationals of symmetric polynomials
\cite{Lee15} which does not have any known analogue for small finite fields. The algorithms obtained for $\kmmd$ and $\kMLC$ are
faster ( $O^{*}(4.08^k)$ time for $\kmmd$ and $O^*(n^{k/2})$ for
$\kMLC$).  In comparison, our algorithms also work for all finite
fields. As already mentioned, the algorithm of Koutis and Williams
\cite{KW16} for $\kMLC$ works over $\F_2$ and the run time is
$O^{*}(n^{k/2})$. In this sense, our algorithm for $\kMLC$ can also be
viewed as a generalization that does not depend on the characteristic
of the ground field. It is to be noted that, over fields of small
characteristic a Waring decomposition of the input polynomial may not
be available. For example, over $\F_2$ the polynomial $xy$ has no
Waring decomposition.

\vspace{0.2 cm}
\noindent\textbf{Organization.}  The paper is organized as follows. In
Section~\ref{section-framework} we explain the Hadamard product
framework. The proof of Theorem \ref{kmlc-thm} and its consequences
are given in Section~\ref{open-problem}. Section~\ref{kmmd} contains
the the proof of Theorem \ref{kmmd-thm}. The proof of Theorem
\ref{depth3-kmmd} can be found in the full version in ArXiv.

\section{Hadamard Product Framework} \label{section-framework}

Computing the Hadamard product of two commutative polynomials is, in
general, computationally hard. This can be observed from the fact that
the Hadamard product of the determinant polynomial with itself is the
permanent polynomial.  Nevertheless, we develop a method for some
special cases, that is efficient with degree $k$ as the fixed
parameter, for computing the \emph{scaled} Hadamard product of
commutative polynomials.

\begin{definition}\label{s-hadprod} 
  The \emph{scaled} Hadamard product of polynomials $f,g \in
  \F[X]$ is defined as
\[
f\circ^{s} g = \sum_{m} (m! \cdot [m]f \cdot [m]g) \ m,
\]
where for monomial $m=x^{e_1}_{i_1}x^{e_2}_{i_2} \ldots x^{e_r}_{i_r}$
we define $m!  = e_1!\cdot e_2!\cdots e_r!$.
\end{definition}

Computing the scaled Hadamard product is key to our algorithmic
results for $\kmmd$ and $\kMLC$. Broadly, it works as follows: we
transform polynomials $f$ and $g$ to suitable \emph{noncommutative}
polynomials. We compute their (noncommutative) Hadamard product
efficiently~\cite{AJS09,AS18}, and we finally recover the scaled
commutative Hadamard product $f\circ^s g$ (or evaluate it at a desired
point $\vec{a} \in\F^n$).

Suppose $f\in\F[x_1,x_2,\ldots,x_n]$ is a homogeneous degree-$k$
polynomial given by a circuit $C$. We can define its noncommutative version $C^{nc}$ which computes
the noncommutative homogeneous degree-$k$ polynomial
$\hat{f}\in\F\angle{y_1,y_2,\ldots,y_n}$ as follows. 

\begin{definition}\label{noncomm-version}
Given a commutative circuit $C$ computing a polynomial in $\F[x_1,
  x_2, . . . , x_n]$, the noncommutative version of $C$, $C^{nc}$ is
the noncommutative circuit obtained from $C$ by fixing an ordering of
the inputs to each product gate in $C$ and replacing $x_i$ by the
noncommuting variable $y_i : 1\leq i \leq n$.
\end{definition}

Let $X_k$ denote the set of all degree-$k$ monomials over $X$. As
usual, $Y^k$ denotes all degree-$k$ noncommutative monomials (i.e.,
words) over $Y$. Each monomial $m\in X_k$ can appear as different
noncommutative monomials $\hat{m}$ in $\hat{f}$. We use the notation
$\hat{m}\to m$ to denote that $\hat{m}\in Y^k$ will be transformed to
$m\in X_k$ by substituting $x_i$ for $y_i, 1\le i\le n$. Then, we
observe the following, $[m]f = \sum_{\hat{m}\to m}[\hat{m}]\hat{f}.$

The noncommutative circuit $C^{nc}$ is not directly useful for
computing Hadamard product.  However, the following symmetrization
helps. We first explain how permutations $\sigma\in S_k$ act on the
set $Y^k$ of degree-$k$ monomials (and hence, by linearity, act on
homogeneous degree $k$ polynomials).

For each monomial $\hat{m}=y_{i_1}y_{i_2}\cdots y_{i_k}$, the
permutation $\sigma\in S_k$ maps $\hat{m}$ to the monomial
$\hat{m}^{\sigma}$ defined as
$\hat{m}^{\sigma}=y_{i_{\sigma(1)}}y_{i_{\sigma(2)}}\cdots
y_{i_{\sigma(k)}}$. By linearity, $\hat{f}=\sum_{\hat{m}\in
  Y^k}[\hat{m}]\hat{f} \cdot \hat{m}$ is mapped by $\sigma$ to the
polynomial, $\hat{f}^\sigma=\sum_{\hat{m}\in Y^k}[\hat{m}]\hat{f}\cdot
\hat{m}^\sigma.$

The \emph{symmetrized polynomial} of $f$, $f^*$, is degree-$k$ homogeneous
polynomial $f^* = \sum_{\sigma\in S_k} \hat{f}^{\sigma}.$
We now explain the use of symmetrization in
computing the scaled Hadamard product $f\circ^s g$. 

\begin{proposition}\label{comm-to-noncomm}
  For a homogeneous degree-$k$ commutative polynomial
  $f\in\F[X]$ given by circuit $C$, and its
  noncommutative version $C^{nc}$ computing polynomial
  $\hat{f}\in\F\angle{Y}$, consider the
  \emph{symmetrized} noncommutative polynomial
$
f^* = \sum_{\sigma\in S_k} \hat{f}^{\sigma}.
$
Then for each monomial $m\in X_k$ and each word $m'\in Y^k$ such that
$m'\to m$, we have:
$
[m']f^* = m!\cdot [m]f.
$
\end{proposition}
\begin{proof}
Let $f=\sum_{m}[m]f\cdot m$ and
$\hat{f}=\sum_{\hat{m}}[\hat{m}]\hat{f}\cdot \hat{m}$. Notice that $[m]f = \sum_{\hat{m}\to m}[\hat{m}]\hat{f}$.
Now, we write $f^*=\sum_{m'}[m']f^*\cdot m'.$
The group $S_k$ acts on $Y^k$ (degree $k$ words in $Y$) by permuting
the positions. Suppose $m=x_{i_1}^{e_1}\cdots x_{i_q}^{e_q}$ is a type
$\e=(e_1,\ldots,e_q)$ degree $k$ monomial over $X$ and $m'\to
m$. Then, by the \emph{Orbit-Stabilizer lemma} the orbit $O_{m'}$ of
$m'$ has size $\frac{k!}{m!}$. It follows that

\[
[m']f^*=\sum_{\hat{m}\in O_{m'}}m!\cdot [\hat{m}]\hat{f}= m!
\sum_{\hat{m}\to m}[\hat{m}]\hat{f}=m!\cdot [m]f.
\] 
It is important to note that for some $\hat{m}\in Y^k$ such that
$\hat{m}\to m$, even if $[\hat{m}]\hat{f} = 0$ then also $[\hat{m}]f^*
= m!\cdot [m]f$.
\end{proof}

Next, we show how to use Proposition \ref{comm-to-noncomm} 
to compute scaled Hadamard product in the commutative setting 
via noncommutative Hadamard product.  We note
that given a commutative circuit $C$ computing $f$, the
noncommutative polynomial $\hat{f}$ depends on the circuit
structure of $C$. However, $f^*$ depends only on the
polynomial $f$. 

\begin{lemma}\label{com-noncomm-relation}
 Let $C$ be a circuit for a homogeneous degree-$k$ polynomial
 $g\in\F[X]$. For any homogeneous degree-$k$ polynomial $f\in\F[X]$,
 to compute a circuit for $f\circ^{s} g$ efficiently, it suffices to
 compute a circuit for $ f^{*}\circ \hat{g}$ efficiently where
 $\hat{g}$ is the polynomial computed by the noncommutative circuit
 $C^{nc}$. Moreover, given any point $\vec{a}\in\F^n$,
  $(f \circ^s  g)(\vec{a}) =( f^{*} \circ \hat{g})(\vec{a})$.
\end{lemma}

\begin{proof}
We write $f= \sum_m [m]f\cdot m$ and $g =
\sum_{m'} [m']g \cdot m'$, and notice that $f\circ^s g = \sum_m m!\cdot [m]f\cdot [m]g\cdot m. $

Suppose the polynomial computed by $C^{nc}$ is $\hat{g}(Y) = \sum_{m\in X_k}\sum_{\hat{m}\to m}[\hat{m}]\hat{g}\cdot \hat{m}.$
By Proposition~\ref{comm-to-noncomm}, the
noncommutative polynomial $f^*(Y) = \sum_{m\in X_k}\sum_{\hat{m}\to m}m!\cdot [m]f\cdot \hat{m}.$
Hence,
\[
( f^{*} \circ \hat{g})(Y) = \sum_{m\in X_k}\sum_{\hat{m}\to m}m!\cdot [m]f\cdot [\hat{m}]g\cdot \hat{m} = \sum_{m\in X_k}m!\cdot [m]f \sum_{\hat{m}\to m}[\hat{m}]g\cdot \hat{m}.
\] 
Therefore, using any commutative substitution (i.e. by substituting the $Y$ variables by $X$ variables), we get back a commutative circuit for $f \circ^s g$. Moreover,
given a point $\vec{a}\in \F^n$,
\[
( f^{*} \circ \hat{g})(\vec{a}) = \sum_{m\in X_k}m!\cdot [m]f \sum_{\hat{m}\to m}[\hat{m}]g\cdot \hat{m}(\vec{a}) =  \sum_{m\in X_k}m!\cdot [m]f\cdot m(\vec{a}) \sum_{\hat{m}\to m}[\hat{m}]g.
\] 
From the definition, $[m]g = \sum_{\hat{m}\to m}\hat{m}[\hat{g}]$. Hence, $( f^{*} \circ \hat{g})(\vec{a}) = \sum_{m\in X_k}m!\cdot [m]f\cdot m(\vec{a}) [m]g = (f \circ^s g)(\vec{a})$.
\end{proof}


\section{The Sum of Coefficients of Multilinear Monomials} \label{open-problem}
 
In this section we prove Theorem~\ref{kmlc-thm}. As already sketched in Section \ref{intro}, the main conceptual step is to apply the symmetrization trick to reduce the $\kMLC$ problem to evaluating rectangular permanent over a suitable matrix ring. Then we use a result of \cite{BHKK10} to solve the instance of rectangular permanent evaluation problem. As corollaries of our technique, we improve the running time of exact counting of several combinatorial problems studied in \cite{KW16}.

Before we prove the theorem, let us recall the definition of an ABP. 
An \emph{algebraic branching program} (ABP) is a
directed acyclic graph with one in-degree-$0$ vertex called
\emph{source}, and one out-degree-$0$ vertex called \emph{sink}. The
vertex set of the graph is partitioned into layers $0,1,\ldots,\ell$,
with directed edges only between adjacent layers ($i$ to
$i+1$). The source and the sink are at layers zero and $\ell$
respectively. Each edge is labeled by a linear form over variables
$x_1,x_2,\ldots,x_n$. The polynomial computed by the ABP is the sum
over all source-to-sink directed paths of the product of linear forms
that label the edges of the path. An ABP is \emph{homogeneous} if all
edge labels are homogeneous linear forms. ABPs can be defined in both
commutative and noncommutative settings. Equivalently, a homogeneous ABP of width $w$ computing a degree-$k$ polynomial over $X$ can be thought of
as the $(1,w)^{th}$ entry of the product of $w\times w$ matrices $M_1\cdots M_k$ where entries of each $M_i$ are homogeneous linear forms over $X$. By $[x_j]M_i$, we denote the $w\times w$ matrix
over $\F$, such that $(p,q)^{th}$ entry of the matrix,$([x_j]M_i)(p,q) = [x_j](M_i(p,q))$, the coefficient of $x_j$ in the linear form of the $(p,q)^{th}$ entry of $M_i$.

We now define the permanent of a rectangular matrix.
The permanent of a rectangular $k\times n$ matrix $A = (a_{ij})$, with $k\leq n$ is defined as
$
\rper(A) = \sum_{\sigma\in I_{k,n}} \prod_{i=1}^k a_{i,\sigma(i)}
$
where $I_{k,n}$ is the set of all injections from $[k]$ to $[n]$.
Also, we define the noncommutative polynomial $S^*_{n,k}$ as
$
S^*_{n,k}(y_1,y_2,\ldots , y_n) = \sum_{T\subseteq [n],|T| = k}\sum_{\sigma\in S_k} \prod_{i\in T}y_{\sigma(i)}
$
which is the symmetrized version of the elementary symmetric polynomial $S_{n,k}$ as defined in Proposition~\ref{comm-to-noncomm}. Given a set of matrices $M_1, \ldots, M_n$ define the rectangular matrix $A=(a_{i,j})_{i\in[k],j\in[n]}$ such that $a_{i,j} = M_j$. Now we make the following crucial observation.
\begin{observation}\label{obs-rper}
 \[ 
 S^{*}_{n,k}(M_1, \ldots,M_n) = \rper(A).
 \] 
\end{observation}
We use a result from \cite{BHKK10}, that shows that over \emph{any} ring $R$, the permanent of a rectangular $k\times n$ matrix can be evaluated using $O^*(k{n\choose {\downarrow k/2}})$ ring operations. In particular, if $R$ is a matrix ring $M_s(\F)$, the algorithm runs in time  $O(k{n\choose {\downarrow k/2}}\poly(n,s))$. Now we are ready to prove Theorem \ref{kmlc-thm}.

\begin{proof}
Let us first proof a special case of the theorem when the polynomial $f$ is given by an ABP $B$ of width $s$. Notice that, we can compute the sum of the coefficients of the degree-$k$ multilinear terms by evaluating $(f\circ S_{n,k})(\vec{1})$. 
 Now to compute the Hadamard product efficiently, we transfer the problem to the noncommutative domain. 
Let $B^{nc}$ defines the noncommutative version of the commutative ABP $B$ computing the polynomial $f$. 
From Lemma \ref{com-noncomm-relation}, it suffices to compute $(B^{nc}\circ S^{*}_{n,k})(\vec{1})$.  Now, the following lemma reduces this to evaluating $S_{n,k}^*$ over matrix ring. We recall the following result from 
\cite{AS18}. 

 \begin{lemma}(Theorem 2 of \cite{AS10}) \label{main-idea}
 Let $f$ be a homogeneous degree-$k$ noncommutative polynomial in $\F\langle{Y\rangle}$ and $B$ be an ABP of width $w$ computing a 
 homogeneous degree-$k$ polynomial $g = (M_1\cdots M_k)(1,w)$  in $\F\langle{Y\rangle}$. Then $(f\circ g)(\vec{1}) = (f({A^B_1}, \ldots,{A^B_n}))(1,(k+1)w)$ 
 where for each $i\in [n]$, ${A^B_i}$ is the following $(k+1)w\times (k+1)w$ block superdiagonal matrix,
 
\[
{A^B_i} =  
 \begin{bmatrix}
 0 &[y_i]M_1 &0 &\ldots &0\\
 0 &0       &[y_i]M_2 &\ldots &0\\
 \vdots &\vdots &\ddots &\ddots &\vdots\\
 0 &0 &0 &\ldots &[y_i]M_k\\
 0 &0 &0 &\ldots &0
\end{bmatrix}.    
\]
 \end{lemma} 

To see the proof, for any monomial $m = y_{i_1}y_{i_2}\cdots y_{i_k}\in Y^k$, 
\[ 
(A^B_{i_1}A^B_{i_2}\cdots A^B_{i_k})(1,(k+1)w)=([y_{i_1}]M_1\cdot [y_{i_2}]M_2\cdots [y_{i_k}]M_k)(1,w)=[m]g,
\]
 from the definition. Hence, we have,
\begin{align*}
  f(A^B_1,A^B_2,\ldots, A^B_n)(1,(k+1)w) &= \sum_{m\in Y^k} [m]f\cdot m(A^B_{i_1},A^B_{i_2},\ldots ,A^B_{i_k})(1,(k+1)w) \\
  &=  \sum_{m\in Y^k} [m]f\cdot [m]g.
\end{align*}

Now, we construct a $k\times n$ rectangular matrix $A = (a_{i,j})_{i\in[k],j\in[n]}$ from the given ABP $B^{nc}$ setting $a_{i,j} = A^{B^{nc}}_j$ as defined.
Using Observation~\ref{obs-rper}, we now have,
\[
\rper(A)(1,(k+1)s) = S^*_{n,k}(A^{B^{nc}}_1,\ldots,A^{B^{nc}}_n)(1,(k+1)s) =  (S^*_{n,k} \circ B^{nc})(\vec{1}) = (S_{n,k}\circ^s B)(\vec{1}).
\] 
Hence combining the algorithm of Bj\"{o}rklund et al.\ for evaluating rectangular permanent over noncommutative ring \cite{BHKK10} with Lemma~\ref{main-idea}, we can evaluate the sum of the coefficients deterministically in time $O(k{n\choose {\downarrow k/2}}\poly(s,n))$. 

Now, we are ready to prove the general case. It uses
the following standard transformation from circuit to ABP ~\cite{VSBR83, SY10} and reduces the problem to the ABP case again.
Given an arithmetic circuit of size $s'$ computing a polynomial $f$ of degree $k$, $f$ can also be computed by a homogeneous ABP of size $s'^{O(\log k)}$.  
Hence given a polynomial $f$  by a $\poly(n)$ sized circuit, we first get a circuit of $\poly(n)$ size for degree-$k$ part of $f$ using standard method of homogenization \cite{SY10}. Then we convert the circuit to a homogeneous ABP of size $n^{O(\log k)}$. 
Then, we apply the first part of the proof on the newly constructed ABP. 
Notice that the entire computation can be done in deterministic $O^*(n^{k/2+c\log k})$ for some constant $c$. 
\end{proof}

\subsection*{Some Applications}
As immediate consequence of Theorem \ref{kmlc-thm}, we improve the counting complexity of several combinatorial problems studied in \cite{KW16}. 
To the best of our knowledge, nothing better than the trivial exhaustive search algorithm were known for the counting version of these problems .
We start with the $\ktreenew$ problem. 

\begin{corollary}
Given a tree $T$ of $k$ nodes and a graph $G$ of $n$ nodes, we can count the number of (not necessarily induced) copies of $T$ in $G$ in deterministic $O^*(\binom{n}{\downarrow k/2})$ time.
\end{corollary}
\begin{proof}Let us define
$Q = \sum_{j\in V(T),i\in V(G)} C_{T,i,j}$, following \cite{KW16}
where 
if $|V(T)|=1$, we define $C_{T,i,j} = x_j$ and
if $|V(T)|>1$, let $T_{i,1},\ldots,T_{i,\ell}$ be the connected subtrees of $T$ remaining after node$i$ is removed from $T$. For each $t\in [\ell]$, let $n_{i,t} \in [k]$ be the (unique) node in $T_{i,t}$ that is a neighbour of $i$ in $T$, then we define
\[
C_{T,i,j} = \prod_{t=1}^\ell \left( \sum_{j':(j,j')\in E(G)} x_j \cdot C_{T_{i,t},n_{i,t},j'} \right).
\]
By the result of \cite{KW16}, it is known that to solve the $\ktreenew$ problem it is sufficient to count the number of multilinear terms in $Q$. 
Following Theorem~\ref{kmlc-thm}, it suffices to show that $Q$ has a $\poly(n,k)$ sized ABP.
It is enough to show that $C_{T,i,j}$ has a $\poly(n,k)$ sized ABP and the ABP for $Q$ follows easily.
We construct an ABP for each $C(T,i,j)$ of size $\poly(n,k)$ by induction on size of $T$. Suppose $C_{T,i,j}$ has such small ABP for $|V(T)| \leq p$. Then, for $V(T) = p+1$,  it is clear from the definition that $C(T,i,j)$ will also have a small ABP. 
Therefore, the polynomial $Q$ will also have an ABP of size $\poly(n,k)$.
\end{proof}

The second application is for $\tdominatingset$ problem. 

\begin{corollary}
Given a graph $G = (V, E)$, we can count the number of sets $S$ of size $k$ that dominates at least $t$ nodes in $G$ in $O^*(\binom {n}{\downarrow t/2})$ deterministic time.
\end{corollary}
\begin{proof}
Following \cite{KW16}, define
\[
P(X,z) = \left( \sum_{i\in V} \left( (1+zx_i) \prod_{j:(i,j)\in E} (1+z x_j)\right) \right)^k.
\]
We inspect $[z^t]P(X,z)$ which is a homogeneous degree $t$ polynomial over $X$, call it $Q(X)$. As $P(X,z)$ has a small ABP of $\poly(n,k)$ size substituing $z$ by any scalar,  we obtain an ABP of size $\poly(n,k)$ for $Q(X)$ also by interpolation. Then, we use the standard method to homogenize the ABP and apply Theorem~\ref{kmlc-thm} to count the number of multilinear terms. This is sufficient to solve the problem by the result of \cite{KW16}. 
\end{proof}
 
The final application is regarding $\mdimensionalkmatching$ problem. 
\begin{corollary}
Given mutually disjoint sets $U_i$, $i\in [m]$, and a collection $C$ of $m$-tuples from $U_1\times \cdots \times U_m$ , we can count the number of sub-collection of $k$ mutually disjoint $m$-tuples in $C$ in deterministic $O^*(\binom{n}{\downarrow (m-1)k/2})$ time.
\end{corollary}
\begin{proof}
Following \cite{KW16} ,
 encode each element $u$ in $U =  \cup_{i=2}^m U_i$ by a variable $x_u \in X$. Encode each $m$-tuple $t = (u_1,\ldots,u_m) \in C\subseteq U_1 \times \cdots \times U_m$ by the monomial $M_t = \prod_{i=2}^m x_{u_i}$. Assume $U_1 = \{u_{1,1},\ldots,u_{1,n}\}$, and let $T_j \subseteq C$ denote the subset of $m$-tuples whose first coordinate is $u_{1,j}$. Consider the polynomial
\[
P(X,z) = \prod_{j=1}^n \left( 1+ \sum_{t\in T_j} (z\cdot M_t)\right).
\]
Clearly, $P(X,z)$ has an ABP of size $\poly(n,m)$. Let $Q(X) = [z^k]P(X,z)$, we can obtain a small ABP of size $\poly(n,m,k)$ for $Q(X)$ by interpolation. Now, we homogenize the ABP and apply Theorem~\ref{kmlc-thm} to count the number of multilinear terms which is sufficient by the result of \cite{KW16}. 
\end{proof}

\subsection*{Hardness for Computing Rectangular Permanent over \emph{any} Ring}

 In \cite{BHKK10}, it is shown that a
$k\times n$ rectangular permanent can be evaluated over commutative
rings and commutative semirings in $O(h(k)\cdot\poly(n,k))$ time for some computable function $h$ . 
In other words, the problem is in FPT parameterized by the number of rows. An
interesting question is to ask whether one can get any FPT
algorithm when the entries are from noncommutative rings (in
particular, matrix rings). We prove that such an algorithm is
unlikely to exist.
We show that counting the number of $k$-paths
in a graph $G$, a well-known $\#\w[1]$-complete problem, reduces to
this problem. So, unless ETH fails we do not have such an algorithm
\cite{DF13}.

\begin{theorem}\label{comp_rect_perm}
Given a $k \times n$ matrix $X$ with entries $X_{ij} \in \mathbb{M}_{t \times t} (\Q)$, 
computing the rectangular permanent of $X$ is $\# \w[1]$-hard with $k$ as the parameter where $t=(k+1)n$ under polynomial time many-one reductions.
\end{theorem}

\begin{proof}
If we have an algorithm to compute the permanent of a $k\times n$
matrix over noncommutative rings which is FPT in parameter $k$,
that yields an algorithm which is FPT in $k$ for evaluating the polynomial $S^*_{n,k}$ on
matrix inputs. This follows from Observation \ref{obs-rper}. Now, given a graph $G$ we can compute a homogeneous $\ABP$ of width $n$ and $k$ layers for the graph polynomial $C_G$ defined as follows.
Let $G(V,E)$ be a
directed graph with $n$ vertices where $V(G)= \{v_1,v_2,\ldots,
v_n\}$.  A $k$-walk is a sequence of $k$ vertices $v_{i_1},
v_{i_2},\ldots, v_{i_k}$ where $(v_{i_j},v_{i_{j+1}})\in E$ for each
$1\leq j \leq k-1$. A $k$-path is a $k$-walk where no vertex is
repeated.  Let $A$ be the adjacency matrix of $G$, and let
$y_1,y_2,\ldots,y_n$ be noncommuting variables. Define an $n\times n$
matrix $B$
\[
B[i, j] = A[i, j]\cdot y_i,~~ 1\leq i,j \leq n.
\]
Let $\vec{1}$ denote the all $1$'s vector of length $n$. Let $\vec{y}$
be the length $n$ vector defined by $\vec{y}[i] = y_i$. The
\emph{graph polynomial} $C_G\in\F\angle{Y}$ is defined as
\[
C_G(Y) = \vec{1}^T\cdot B^{k-1}\cdot \vec{y}.
\]
Let $W$ be the set of all $k$-walks in $G$. The following observation
is folklore.

\begin{observation} \label{obs-graph-poly}
\[
C_G(Y) =  \sum_{v_{i_1}v_{i_2}\ldots v_{i_k}\in W} y_{i_1}y_{i_2}\cdots y_{i_k}.	
\]
Hence, $G$ contains a $k$-path if and only if the graph polynomial
$C_G$ contains a multilinear term.
\end{observation}

Clearly the number of $k$-paths in $G$ is equal to $(C_G\circ S_{n,k})(\vec{1})$. By Lemma \ref{com-noncomm-relation}, we know that it suffices to compute  $({C_G}^{nc}\circ S^{*}_{n,k})(\vec{1})$. We construct $kn \times kn$ matrices $A_1, \ldots, A_n$ from the ABP of ${C_G}^{nc}$ following Lemma~\ref{main-idea}. Then  
from Lemma~\ref{main-idea}, we know that $({C_G}^{nc}\circ S^{*}_{n,k})(\vec{1})=S^{*}_{n,k}(A_1, \ldots, A_n)(1,t)$ where $t=(k+1)n$. 
So if we have an algorithm which is FPT in $k$ for evaluating $S^{*}_{n,k}$ over matrix inputs, we also get an algorithm to count the number of $k$-paths in $G$ in FPT$(k)$ time. 
\end{proof}

\section{Multilinear Monomial Detection} \label{kmmd}

In this section, we prove Theorem~\ref{kmmd-thm}. Apart from being a new technique, the Hadamard product based algorithm runs in polynomial space and does not depend on the characteristic of the ground field. This is in contrast with the exterior algebra based approach \cite{BDH18} and Waring rank based approach \cite{Pra18}.    

We first recall that
the Hadamard product of a noncommutative circuit and a noncommutative
ABP can be computed efficiently. 
The proof is similar to the proof of Lemma~\ref{main-idea}.

\begin{lemma}(Corollary 4 of \cite{AS18}) \label{abp-circuit}
Given a homogeneous noncommutative circuit of size $S'$ for $f\in
\F\angle{y_1,y_2,\ldots,y_n}$ and a homogeneous noncommutative ABP of
size $S$ for $g\in \F\angle{y_1,y_2,\ldots,y_n}$, we can compute a
noncommutative circuit of size $O( S^3S^\prime)$ for $f\circ g$ in
deterministic $S^3S'\cdot \poly(n,k)$ time where $k$ is the degree
upper bound for $f$ and $g$.
\end{lemma}
 Now we give an algorithm for computing the Hadamard product 
for a special case in the commutative setting. Any depth two $\Pi^{[k]}\Sigma$ circuit 
computes the product of $k$ homogeneous linear forms over the input set of variables 
$X$.

\begin{lemma}\label{depth2-Hadamard}
Given an arithmetic circuit ${C}$ of size $s$ computing $g\in
\mathbb{F}[X]$, and a homogeneous $\Pi^{[k]}\Sigma$ circuit computing
$f\in \F[X]$, and any point $\vec{a}\in\F^n$, we can evaluate
$(f\circ^s g)(\vec{a})$ in $O^*(2^k)$ time and in polynomial space.
\end{lemma}

\begin{proof}
By standard homogenization technique \cite{SY10} we can extract the homogeneous degree-$k$ component of $C$ and thus we can assume that $C$ computes a homogeneous
degree-$k$ polynomial. Write $f=\prod_{j=1}^k L_{j}$, for homogeneous
linear forms $L_{j}$. The corresponding noncommutative polynomial
$\hat{f}$ is defined by the natural order of the $j$ indices (and
replacing $x_i$ by $y_i$ for each $i$).  
\begin{claim}\label{depth2symm}
The noncommutative polynomial $f^*$ has a (noncommutative)
$\Sigma^{[2^k]}\Pi^{[k]}\Sigma$ circuit, which we can write as
$f^*=\sum_{i=1}^{2^k}C_i$, 
where each $C_i$ is a (noncommutative) $\Pi^{[k]}\Sigma$ circuit. 
\end{claim}

Before we prove the claim, we show that it easily yields the desired
algorithm: First we notice that
\[
C^{nc}\circ f^{*} = \sum_{i=1}^{2^k}C^{nc}\circ C_i.
\]
Now, by Lemma~\ref{abp-circuit}, we can
compute in $\poly(n,s,k)$ time a $\poly(n,s,k)$ size circuit
for the (noncommutative) Hadamard product $C^{nc}\circ
C_i$. As argued in the proof of Lemma~\ref{com-noncomm-relation}, 
for any $\vec{a}\in\F^n$ we have 
\[
(g\circ^s f)(\vec{a}) = (C\circ^s f)(\vec{a}) = (C^{nc}\circ f^*)(\vec{a}).
\] 
Thus, we can evaluate $(g\circ^s f)(\vec{a})$ by incrementally computing
$(C^{nc}\circ C_i)(\vec{a})$ and adding up for $1\le i\le 2^k$. This can be clearly 
implemented using only polynomial space. 
\end{proof}

Now, we prove the above claim. Consider $f=L_{1}L_{2}\cdots
L_{k}$. Then $\hat{f}=\hat{L}_1 \hat{L}_{2}\cdots \hat{L}_{k}$, where
$\hat{L}_{j}$ is obtained from $L_{j}$ by replacing variables $x_i$
with the noncommutative variable $y_i$ for each $i$.  We will require
the following observation.

\begin{observation}\label{important}
\[
f^* = \sum_{\sigma\in S_k}
\hat{L}_{\sigma(1)}\hat{L}_{\sigma(2)}\cdots \hat{L}_{\sigma(k)}.
\]
\end{observation}
\begin{proof}
Let us prove the claim, monomial by monomial. Fix a monomial $m'$ in
$f^*$ such that $m'\to m$. Suppose $m' = y_{i_1}y_{i_2}\ldots
y_{i_k}$. Note that, $m = x_{i_1}x_{i_2}\ldots x_{i_k}$. Recall from
Proposition~\ref{comm-to-noncomm}, $[m']f^* = m!\cdot [m]f$. Now, the
coefficient of $m'$ in $ \sum_{\sigma\in S_k} \prod_{j=1}^k
\hat{L}_{\sigma(j)}$ is
\[
[m']\left( \sum_{\sigma\in S_k} \prod_{j=1}^k \hat{L}_{\sigma(j)} \right) = \sum_{\sigma\in S_k} \prod_{j=1}^k [y_{i_j}] \hat{L}_{\sigma(j)}. 
\]
Let us notice that, $[y_{i_j}] \hat{L}_{\sigma(j)} = [x_{i_j}]
L_{\sigma(j)}$. Hence,
\[
[m']\left( \sum_{\sigma\in S_k} \prod_{j=1}^k \hat{L}_{\sigma(j)} \right) = \sum_{\sigma\in S_k} \prod_{j=1}^k [x_{i_j}] L_{\sigma(j)}.
\]
\end{proof} 

Now we observe the following easy fact. 
\begin{observation}\label{calc}
 For a degree $k$ monomial $m = x_{i_1}x_{i_2}\cdots x_{i_k}$ (where
 the variables can have repeated occurrences) and a homogeneous
 $\Pi^{[k]}\Sigma$ circuit $C=\prod^k_{j=1} L_j$, the coefficient of monomial
 $m$ in $C$ is given by
 $
  m!\cdot [m]C = \sum_{\sigma \in S_k} \prod^k_{j=1}( [x_{i_j}]
  L_{\sigma(j)}).
 $
\end{observation}

\begin{tpproof}
Now, the claim directly follows from Observation~\ref{calc} as
$\sum_{\sigma\in S_k} \prod_{j=1}^k [x_{i_j}] L_{\sigma(j)} = m!\cdot
[m]f$. 

Now define the $k\times k$ matrix $T$ such that each row of $T_i$ is
just the linear forms $\hat{L}_{1}, \hat{L}_{2}, \ldots , \hat{L}_{k}$
appearing in $f$. The (noncommutative) permanent of $T$ is given by
$\perm(T)=\sum_{\sigma \in S_k} \prod^k_{j=1} \hat{L}_{\sigma(j)}$,
which is just $f^*$.

We now apply Ryser's formula \cite{ry63} (noting the
fact that it holds for the noncommutative permanent too), to express
$\perm(T)$ as a depth-3 homogeneous noncommutative
$\Sigma^{[2^k]}\Pi^{[k]}\Sigma$ formula. It follows that
$f^*=\perm(T)$ has a $\Sigma^{[2^k]}\Pi^{[k]}\Sigma$ noncommutative
formula.
\end{tpproof}\qed

We include a proof of Observation \ref{calc} for completeness. 

\begin{proof}
We assume, without loss of generality, that repeated variables are
adjacent in the monomial $m=x_{i_1}x_{i_2}\cdots x_{i_k}$. More
precisely, suppose the first $e_1$ variables are $x_{j_1}$, and the
next $e_2$ variables are $x_{j_2}$ and so on until the last $e_q$
variables are $x_{j_q}$, where the $q$ variables $x_{j_k}, 1\le k\le
q$ are all distinct.

We notice that the monomial $m$ can be generated in $C$ by first fixing
an order $\sigma : [k] \mapsto [k]$ for multiplying the $k$ linear
forms as $L_{\sigma(1)}L_{\sigma(2)}\cdots L_{\sigma(k)}$, and then
multiplying the coefficients of variable $x_{i_j}, 1\le j\le k$ picked
successively from linear forms $L_{\sigma(j)}, 1\leq j\leq k$. However,
these $k!$ orderings repeatedly count terms. 

We claim that each distinct product of coefficients term is counted
exactly $m!$ times. Let $E_j\subseteq [k]$ denote the interval
$E_j=\{\ell\mid e_{j-1}+1\le \ell\le e_j\}, 1\le j\le q$, where we set
$e_0=0$.

Now, to see the claim we only need to note that two permutations
$\sigma,\tau\in S_k$ give rise to the same product of coefficients
term if and only if $\sigma(E_j)=\tau(E_j), 1\le j\le q$. Thus, the number
of permutations $\tau$ that generate the same term as $\sigma$
is $m!$. 

Therefore the sum of 
product of coefficients $\sum_{\sigma \in S_k}
\prod^k_{j=1}( [x_{i_j}] L_{\sigma(j)})$ is same as $m!\cdot [m]C$,  which completes the proof.
\end{proof}

\begin{remark}
Over rationals, computing $f\circ^{s}g$, when $g$ is a
$\Pi^{[k]}\Sigma$ circuit, can also be done by computing $g^*$
using Fischer's identity \cite{Fis94}. However,
Lemma~\ref{depth2-Hadamard} also works over finite fields.
\end{remark}

\vspace{0.2cm}
Now we are ready to prove Theorem \ref{kmmd-thm}.
\begin{proof}
By homogenization, we can assume that $C$ computes a
homogeneous degree $k$ polynomial $f$.

We go over a collection of colorings $\{\zeta_i:[n]\to[k]\}$ chosen
uniformly at random and define a $\Pi^{[k]}\Sigma$ formula $ P_i =
\prod_{j=1}^k \sum_{\ell : \zeta_i(\ell) = j} x_\ell $ for each
colouring $\zeta_i$. A monomial is \emph{covered} by a coloring
$\zeta_i$ if the monomial is nonzero in $P_i$. The probability that a
random coloring covers a given degree-$k$ multilinear monomial is
$\frac{k!}{k^k} \approx e^{-k}$. Hence, for a collection of $O^{*}(e^k)$
many colorings $\{\zeta_i:[n]\to[k]\}$ chosen uniformly at random,
with constant probability all the multilinear terms of degree $k$ are
covered.

For each coloring $\zeta_i$, we construct a circuit $C'_i=C\circ^s
P_i$. 

Notice that we
are interested only in multilinear monomials and for each such
monomial $m$, the additional multiplicative factor $m!=1$. Also, the
coefficient of each monomial is exactly $1$ in each $P_i$, and if $f$
contains a multilinear term then it will be covered by \emph{some}
$P_i$.  Now we perform $\pit$ test on each $C_i'$ using
Demillo-Lipton-Schwartz-Zippel Lemma~\cite{DL78, Zip79, Sch80} in
randomized polynomial time to complete the procedure. More precisely, we pick a random $\vec{a}\in \F^n$ and evaluate $C'_i$ on that point. Notice that, by the proof of Lemma~\ref{depth2-Hadamard}, it is easy to see that $C_i'(\vec{a})$ can be computed deterministically in time $2^k\cdot\poly(n,s)$ time and $\poly(n,k)$ space. \footnote{
Since the syntactic degree of the circuit is not bounded here, and if we have
to account for the bit level complexity (over $\Z$) of the scalars generated in 
the intermediate stage we may get field elements whose bit level complexity is exponential 
in the input size. So, a standard technique is to take a random prime of polynomial bit-size 
and evaluate the circuit modulo that prime.} 

To improve the run time from $O^*((2e)^k)$ to $O^*(4.32^k)$, we can
use the idea of H\"{u}ffner et al.\cite{HWZ08}\footnote{This is also
 used in \cite{BDH18}.}. The key idea is that, using more than $k$
colors we would reduce the number of colorings and hence the number of
$\Pi\Sigma$ circuits, but it would increase the formal degree of each
$P_i$. Following \cite{HWZ08}, we use $1.3 k$ many colors and each
$P_i$ will be a $\Pi^{[1.3k]}\Sigma$ circuit. For each coloring
$\zeta_i:[n]\to [1.3k]$ chosen uniformly at random, we define the
following $\Pi^{[1.3k]}\Sigma$ circuit, $ P_i(x_1,x_2,\ldots ,x_n,z_1,
\ldots, z_{1.3k}) = \prod_{j=1}^{1.3k} \left( \sum_{\ell :
  \zeta_i(\ell) = j} x_\ell +z_j\right).  $

Since each $P_i$ is of degree $1.3k$, we need to modify the circuit
$C$ to another circuit $C'$ of degree $1.3k$ in order to apply
Hadamard products. The key idea is to define the circuits
$C'\in\F[X,Z]$ as follows:
\[
C'(X,Z)= C(X)\cdot S_{1.3k,0.3k}(z_1,\ldots,z_{1.3k})
\] 
where $S_{1.3k,0.3k}(z_1,\ldots,z_{1.3k})$ is the elementary symmetric
polynomial of degree $0.3k$ over the variables $z_1, \ldots,
z_{1.3k}$. 
By the result of \cite{HWZ08}, for
$O^{*}(1.752^k)$ many random colorings with high probability each
multilinear monomial in $C$ will be covered by the monomials of some
$P_i$ (over the $X$ variables).

Now to compute $C'^{nc}\circ P_i^{*}$ for each $i$, we symmetrize the polynomial $P_i$, the symmetrization happens
over the $X$ variables as well as over the $Z$ variables. But in
$C'^{nc}$ we are only interested in the monomials (or words) where the
rightmost $0.3k$ variables are over $Z$ variables. In the
noncommutative circuit $C'^{nc}$, every sub-word $z_{i_1} z_{i_2}
\ldots z_{i_{0.3k}}$ receives a natural ordering $i_1 <
i_2<\ldots<i_{0.3k}$.

Notice that 
\[
P^{*}_i(X, Z) = \sum_{\sigma\in S_{1.3k}} \prod_{j=1}^{1.3k}
\left( \sum_{\ell : \zeta_i(\ell) = \sigma(j)} x_\ell
+z_{\sigma(j)}\right).
\]

Our goal is to understand the part of $P^{*}_i(X,Z)$ where each
monomial ends with the sub-word $z_{i_1} z_{i_2} \ldots z_{i_{0.3k}}$
and the top $k$ symbols are over the $X$ variables.  For a fixed set
of indices $W=\{i_1 < i_2<\ldots<i_{0.3k}\}$, define the set
$T=[1.3k]\setminus W$.  Let $S_{[k],T}$ be the set of permutations
$\sigma\in S_{1.3k}$ such that $\sigma : [k]\rightarrow T$ and
$\sigma(k+j) = i_j$ for $1\leq j\leq 0.3k$.
As we have fixed the last $0.3k$ positions, each $\sigma\in S_{[k],T}$ corresponds to some $\sigma'\in S_k$.
Let $Z_{W} = z_{i_1} z_{i_2} \ldots z_{i_{0.3k}}$. 
Now the following claim
is immediate.

\begin{claim}\label{part-hadamard}
The part of $P^{*}_i(X,Z)$ where each monomial ends with the
sub-word $Z_W$ and the first $k$
variables are from $X$, is $P^{*}_{i,W} \cdot Z_W$, where
\[
P^{*}_{i,W} (X)= \sum_{\sigma\in S_{[k], T}} \prod_{j=1}^k \left( \sum_{\ell : \zeta_i(\ell) = \sigma(j)} x_\ell \right)
=\sum_{\sigma'\in S_k} \prod_{j=1}^k \left( \sum_{\ell : \zeta_i(\ell) = \sigma'(j)} x_\ell \right).
\]
\end{claim} 

Notice that, $\sum_{W\subseteq [1.3k] : |W|=0.3k} P^*_{i,W}$ contains
all the \emph{colourful} degree-$k$ multilinear monomials over $X$.  We now obtain
the following. 
\[
(C'^{nc} \circ P^{*}_i)(X,Z) = \sum_{W\subseteq [1.3k] : |W|=0.3k}
\left(C^{nc}(X) \circ P^{*}_{i,W}(X)\right)\cdot Z_W.
\]
Setting each $z_i = 1$ and using distributivity of Hadamard product, we get 
$(C'^{nc} \circ P^{*}_i)(X,\vec{1}) = C^{nc}(X) \circ \sum_{W\subseteq [1.3k] : |W|=0.3k} P^{*}_{i,W}$ which is the \emph{colourful} multilinear
part of the input circuit.

We now consider
$(C'\circ^{s} P_i)(X,Z)$ and
substitute $1$ for each $Z$ variable and do a randomized PIT test on
the $X$ variables using Demillo-Lipton-Schwartz-Zippel
Lemma~\cite{DL78, Zip79, Sch80}.
By Lemma~\ref{depth2-Hadamard}, for any random $\vec{a}\in \F^n$, $(C'\circ^{s} P_i)(\vec{a})$ can be computed
in $O^{*}(2^{1.3k}) = O^{*}(2.46^k)$ time and $\poly(n,k)$ space. 
 This suffices to check whether the
resulting circuit is identically zero or not. We repeat the procedure
for each coloring and obtain a randomized $O^*(4.32^k)$ algorithm. This completes the proof of Theorem~\ref{kmmd-thm}.
\end{proof}

\vspace{2mm} 
\noindent\textbf{Acknowledgement} We thank anonymous reviewers for their comments on an earlier version of this paper. We are particularly grateful to an anonymous reviewer for pointing out the combinatorial applications of Theorem \ref{kmlc-thm} in exact counting. 
 
\newpage
\bibliography{ref2}

\end{document}